\documentclass[fleqn,twoside]{article}%
\usepackage{amsmath}
\usepackage{amsfonts}
\usepackage{amssymb}
\usepackage{graphicx}%
\def\be{\begin{equation}}

\def\ee{\end{equation}}
\def\bes{\begin{equation}\begin{split}&}
\def\es{\end{split}}
\def\bi{\bibitem}

\begin{document}
\title{Extended Inflation - a quantum cosmological survey.}
\author{Abhik Kumar Sanyal$^\dag$ and Bijan Modak $^\ddag$}

\maketitle
\noindent
\begin{center}
\noindent
$^{\dag}$ Dept. of Physics, Jangipur College, Murshidabad, West Bengal, India-742213\\
\noindent
$^{\ddag}$ Dept. of Physics, University of Kalyani, West Bengal, India - 741235.\\

\end{center}
\footnotetext[1] {\noindent
Electronic address:\\
\noindent
$^{\dag}$ sanyal\_ak@yahoo.com\\
$^{\ddag}$ bmodak@yahoo.com}

\begin{abstract}
This paper considers Einstein-Brans-Dicke action being coupled with a Higgs sector. It is shown that classically graceful exit from inflation is not possible in principle, rejecting the claim of La and Steinhardt. Further, wormhole solutions with and without conserved charges are explored. Finally, the wavefunction of the universe in this model is constructed with the boundary condition, as proposed by Vilenkin. The interesting feature of the wavefunction is that, it has been constructed for an arbitrary factor-ordering index.
\end{abstract}

\section{Introduction}
In the last decade, scientists all over the world have initiated lot of research activities on the semiclassical treatment of gravity namely quantum cosmology, since there has so far been no complete and consistent quantum theory of gravity. This arose mainly because of the interesting features of the models of the ``inflationary universe" \cite{1,2,3}. Usually  quantum fluctuations have been believed to lead to classical density perturbations \cite{4,5}. However, there should exists a correspondence between quantum fluctuations and classical stochastic averages, which has been introduced in an ad-hoc manner in the inflationary models. Further, the fact that there exists no mechanism for a graceful exit from inflation, is a great obstacle for present day physicists. This is the so called cosmological constant problem in disguise. Along with the aforementioned problems, one faces the greatest difficulty, which is the instability to gravitational collapse.\\

Presently, wormholes are supposed to be a way out of these problems. Wheeler \cite{6} argued long ago that at a distance of the Planck scale, space-time geometry would fluctuate, i.e. there would be a change in the topology of space-time. Recently, Baum \cite{7}, Hawking \cite{8} and Coleman \cite{9} have argued that wormholes might be responsible for the vanishing of the cosmological constant. This interesting issue is presently under debate \cite{10,11,12}. Motivation to work in the field of wormhole physics was initiated by Giddings and Strominger \cite{13}, who were the first to give a complete wormhole solution. The usual wormhole solution is accepted as representing tunnelling between spaces with different topology through an  evolving / non-evolving throat of finite radius, and hence one gets rid of the classical gravitational  collapse.\\

For the existence of wormhole solutions, the Euclidean Ricci tensor must have negative eigenvalues at some point. This property holds for a third rank antisymmetric tensor field in general,. However, it has recently been argued \cite{14,15} that the scalar field wormhole solutions indeed exists, because the scalar field in the Euclidean section need not be real. Further, La and Steinhardt \cite{16} recently claimed that a graceful exit from inflation could be achieved through the introduction of a Brans-Dicke field in the Einstein action with a cosmological constant term for a flat FRW model. This paper investigates the possibility of such an exit for a closed FRW model. Further investigation has been carried out into the nature of the wormhole solutions for such a model.\\

This paper is organized as follows. In section 2, we shall present field equations for the Brans-Dicke scalar-tensor theory minimally coupled to a Higgs field and some immediate consequences of the field equations. In section 3, we shall have a discussion with classical solutions at different regimes and indicate that a graceful exit from inflation is not plausible through the introduction of a Brans-Dicke field. In section 4, we shall carry out a discussion with wormholes in the background of a Brans-Dicke field coupled with a Higgs field. In the absence of a Higgs field, wormhole solutions are the same as that obtained by Giddings and Strrominger, while the two asymptotic de-Sitter regions are connected by a throat in the presence of Higgs field. Section 5 will be devoted to the solution of the Wheeler-DeWitt equation, following Vilenkin's proposal \cite{17}. Some of the solutions are presented with arbitrary factor ordering-index.

\section{Field Equations, Wheeler-DeWitt Equation and Some immediate Consequences}

We start with the Jordan-Brans-Dicke \cite{18} action, which is minimally coupled with the Higgs sector as,

\be \label{2.1} S = \int \sqrt{-g} d^4 x\left(\phi R - {\omega\phi_{,\mu}\phi^{,\mu}\over \phi} + {16\pi\over c^4}L_m\right) + \Sigma,\ee
where, $\phi$ is the Jordan-Brans-Dicke field, $\omega$ is a coupling parameter, $\Sigma$ is the supplementary boundary term and $L_m$ is the associated Lagrangian for the Higgs field, given by

\be \label{2.2} L_m = -{1\over 8\pi}\left(\chi_{,\mu}\chi^{,\mu} + V(\chi)\right),\ee
$\chi$ being the Higgs field. Under conformal transformation, the action \eqref{2.1} leads to

\be \label{2.3} S = \int \sqrt{-g} d^4 x\left(R - {1\over 2}(2\omega +3){\lambda_{,\mu}\lambda^{,\mu}\over \lambda^2} + {16\pi G_0\over c^4}L_m\right) + \Sigma,\ee
If we consider a closed Robertson-Walker metric in the form

\be \label{2.4} ds^2 = -dt^2 + a(t)^2 d\Omega_3^2,\ee
$d\Omega_3^2$ being the metric of a unit three-sphere, the action \eqref{2.3} can be recast into the form

\be \label{2.5} S = 2\pi^2 \int\left[-6 a\dot a^2 + \alpha a^3 \dot\phi^2 +\beta a^3\dot\chi^2 + 6a - \beta a^3 V(\chi)\right] dt, \ee
with the choice $\lambda = e^\phi$, $\alpha = {1\over 2}(2\omega+3)$, and $\beta = 2{G_0\over c^4}$. The field equations are,

\be \label{2.6} 2{\ddot a\over a} + {\dot a^2\over a^2} + {1\over a^2} + {\alpha\over 2}\dot\phi^2
+{\beta\over 2}[\dot\chi^2 - V(\chi)] = 0,\ee
\be \label{2.7} \ddot \phi + 3{\dot a\over a} \dot\phi =0,\ee
\be \label{2.8} \ddot\chi + 3 {\dot a\over a}\dot \chi +{1\over 2} {\partial V\over \partial \chi} = 0,\ee
\be \label{2.9} {\dot a^2\over a^2}  +{1\over a^2} - {\alpha\over 6} \dot\phi^2 - {\beta\over 6} [\dot\chi^2 + V(\chi)] =0.\ee
The Wheeler-DeWitt equation $\hat{H} \Psi = 0$,  for the above model takes the form

\be\label{2.10} \left[{\partial^2\over \partial a^2} + {p\over a}{\partial\over \partial a} -{6\over \alpha a^2}{\partial^2\over \partial \phi^2} -{6\over \beta a^2}{\partial^2\over \partial \chi^2} -U(a,\chi)\right]\Psi = 0,\ee
where,

\be \label{2.11} U(a, \chi) = 24 a^2[6-\beta a^2 V(\chi)],\ee
and operator ordering index $p$ resolves some operator ordering ambiguities. Equation \eqref{2.10} represents a three-dimensional minisuperspace model, which can be divided into a classically forbidden region $U >0$, i.e. $a^2 < {6\over \beta V(\chi)}$, and a classical allowed region $U < 0$, i.e. $a^2 > {6\over \beta V(\chi)}$. The boundary of these two regions is $U = 0$, i.e. $a^2 = {6\over \beta V(\chi)}$. We shall study the Wheeler-DeWitt equation and the wave function of the universe in some detail in section 5. At present, we concentrate on the classical solutions.\\

The first integral of equation \eqref{2.7} yields

\be \label{2.12} a^3\dot \phi = l,\ee
$l$ being the constant of integration. Now in the region where semiclassical gravity plays the dominant role, $V(\chi)$ may be assumed to be slowly varying function of $\chi$, i.e.

\be \label{2.13} \left|{1\over V}{ dV\over d\chi}\right| << 1, ~~ V << 1.\ee
So, neglecting $\partial V\over \partial\chi$ in equation \eqref{2.8}, one gets

\be \label{2.14} a^3\dot\chi = k,\ee
$k$ being yet another constant of integration. Hence, equation \eqref{2.9} takes the form

\be \label{2.15} -\dot a^2 -1 + {n^4\over a^4} + {\beta\over 6}a^2 V =0,\ee
where, $n^4 = {\alpha\over 6}l^2 + {\beta\over 6} k^2$. Equation \eqref{2.15} can be viewed as the equation of motion of a particle moving in a potential

\be \label{2.16} V_1 = 1 - {n^4\over a^4} - {\beta\over 6}a^2 V.\ee
This potential diverges both as $a\rightarrow 0$ and $a\rightarrow -\infty$. So the system encounters both short distance instability, - i.e. instability to gravitational collapse, which means unavoidable singularity, and long distance instability - i.e. instability to inflation, which means that the scale factor $``a"$, and hence the universe expands unboundedly \cite{19}. This fact directly contradicts La-Steinhardt's result \cite{16} that graceful exit from inflation is possible with the inclusion of Brans-Dicke field. It will be made more explicit through some particular solutions, which we shall present in the following section. But before that let us find the condition for obtaining bound solutions. Now, in order that a minimum of $``a"$ at some finite time may exists, it should satisfy the conditions $\dot a = 0$ and $\ddot a > 0$. In view of equation \eqref{2.15} with the condition $\dot a = 0$, one obtains

\be \label{2.17} {\beta\over 6} V a^6  = a^4 - n^4.\ee
Further, taking the time derivative of equation \eqref{2.15} or using equation \eqref{2.15} in equation \eqref{2.16} one gets,

\be \label{2.18} \ddot a = -{2n^4\over a^5} + {\beta\over 6} a V.\ee
For $\ddot a > 0$, one therefore requires ${\beta\over 6} a V > {2n^4\over a^5}$. Using equation \eqref{2.17} one therefore finds $a^4 > 3n^4$, or equivalently,

\be \label{2.19} a^2 \ge {4\over \beta V}.\ee
Equation \eqref{2.19} is thus the condition that a minimum of $``a"$ exists. However, our classical allowed region exists for $a^2 \ge {6\over \beta V}$, so not the whole region of bound solution lies within the classically allowed domain. Whatsoever, if we consider $a^2 \gg {4\over \beta V}$, i.e. $a^4 \gg 3n^4$, then solutions will be obtained which represent bound solutions within classically allowed domain.

\section{Classical Solutions In Different Regimes}

In the very early epoch, as $a \rightarrow 0$, $U(a, \chi) \rightarrow 0$. The corresponding classical solutions are obtained from  equations \eqref{2.12}, \eqref{2.14}, and \eqref{2.15} as

\be\label{3.1} a = \pm 3n^2 (t-t_0)^{1\over 3}; ~~~~\phi \approx {l\over 3n^2}\ln{(t-t_0)};~~~~\chi\approx{k\over 3n^2}\ln{(t-t_0)},\ee
where $t_0$ is a constant of integration. In the above plus sign corresponds to the expanding model, while the minus sign corresponds to the contracting model. Next, if we consider $V(\chi)  = 0$, the solution to equation \eqref{2.15} is

\be \label{3.2} \pm {t\over n} = \sqrt{2} E\left[\gamma, {1\over \sqrt 2}\right] - {1\over\sqrt 2}F\left[\gamma, {1\over\sqrt 2}\right].\ee
where, $F\left[\gamma, {1\over\sqrt 2}\right]$ and $E\left[\gamma, {1\over \sqrt 2}\right]$ are elliptical integrals of first and second kind respectively, and $\gamma = \cos^{-1}({a\over n})$. This solution has a finite time singularity, and since it is a solution with zero cosmological constant, it does not lead to exponential expansion.\\

On the other hand, if $V(\chi)$ sits at a maximum value $V_0$ initially, then three cases arises, which correspond to three different epochs. In what follows, ${\beta V_0\over 6}$ has been replaced by $\Lambda$.\\

For $a\ll n$, and ${n^4\over a^4} \gg 1$, neglecting the factor $``1"$ in equation \eqref{2.15}, we obtain the following solution, viz.

\be\label{3.3} a^2 = {n^2\over \Lambda} \sin{h({3\sqrt{\Lambda} t})}.\ee
The above solution implies that the universe enters inflationary phase with a time scale $t \ge {1\over 3\sqrt{\Lambda}}$. We can also look at the solution as one for flat three-space - the case that was studied earlier by La and Steinhardt \cite{16}.\\

However for $a \approx n$, the solution is simply

\be\label{3.4} a = K_1 \exp{(\sqrt{\Lambda}, t)},\ee
where, $K_1$ is a constant of integration. At this stage, the universe undergoes exponential expansion with a time scale $t \ge {1\over\sqrt \Lambda}$. This solution can also be viewed as that for the flat space gravity with a cosmological constant term.\\

Finally, as the universe expands, the scale factor grows, and since $a \gg n$, one can ignore the $n^4\over a^4$ term from equation \eqref{2.15}. The solution then turns out to be

\be \label{3.5} a = {1\over\sqrt \Lambda}\cos{h(\sqrt{\Lambda} t)}.\ee
This solution asymptotically goes over to exponential expansion with a time scale $t > {1\over \sqrt \Lambda}$, and the universe inflates unboundedly. It can also be viewed as the solution for the closed universe with a cosmological constant. Further, it is the bounce solution in the classically allowed region. We might conclude in view of the solution \eqref{3.5} that in spite of the fact that a Brans-Dicke field has been introduced, the solution in the classically allowed domain is independent of the field, which simply discards the claim of La and Steinhardt's result that a graceful exit is possible through the introduction of a Brans-Dicke field \cite{16}.

\section{Wormholes}

In the last section, we noticed that the only classically accepted solution is the one given by \eqref{3.5}. This solution is free from initial big-bang singularity, explains why the universe is isotropic on the large scale at the present epoch, and lies in the classically allowed regime. However, inflation is unbounded and at present, we do not have any mechanism for graceful exit from inflation. Even if we leave this question aside, the other question obviously appears, and that is: ``What happens in the classically forbidden regime and how does one overcome the initial singularity problem in that regime?" Wormholes appear as a rescue. As has been pointed out in the introduction, at a distance of the Planck's scale, which is essentially classically forbidden regime, the space-time geometry fluctuates, i.e. there is a change in topology. Wormholes connect two asymptotic regions by a throat of finite radius, and in this manner, one can mediate spatial topology change due to the production and absorption of baby universes. So in this section we are going to study the possibility of existence of wormhole solutions for the Einstein-Brans-Dicke theory minimally coupled to a Higgs field.\\

Replacing $t$ by $-i\tau$, $\tau$ being the Euclidean time, we see that the ($^0_0$) equation of Einstein \eqref{2.9} takes the form

\be \label{4.1} a_{,\tau}^2 = 1 + {\alpha\over 6} a^2\phi_{,\tau}^2 + {\beta\over 6} a^2\chi_{,\tau}^2 - {\beta\over 6} a^2 V(\chi),\ee
and the field equations \eqref{2.7}  and \eqref {2.8} take the form

\be \label{4.2} \phi_{,\tau\tau} + 3 {a_{,\tau}\over a}\phi_{,\tau} =0,\ee
\be \label{4.3} \chi_{,\tau\tau} + 3 {a_{,\tau}\over a}\chi_{,\tau} = {1\over 2}{\partial V\over \partial \chi}.\ee
Differentiating equation \eqref{4.1} with respect to $\tau$ and using equation \eqref{4.2}, we get

\be \label{4.4} a_{,\tau\tau} = -{\alpha\over 3} a\phi_{,\tau}^2 - {\beta\over 3} a\chi_{,\tau}^2 - {\beta V(\chi)\over 3} a. \ee
Now for wormholes to exist, $``a"$ should start from a large value in the infinite past, evolve down to a throat of finite radius, and then again increase indefinitely. So $a_{,\tau\tau}$ should be positive, which is possible if $\phi$ and/or $\chi$ are imaginary. Question that obviously arises is ``What is really meant by imaginary field variable in the Euclidean regime?" Since we are interested in tunnelling amplitude, we are not sure whether the classical paths should be real. For example, if there is a cyclic variable in the theory, the corresponding Lorentzian momentum would have been a real conserved quantity and not the Euclidean one. In that case the square of the Euclidean momentum is negative, which means that the field variable is imaginary. For a massless scalar field, Brown et al \cite{14} have shown that the Euclidean configuration matches smoothly at the throat to a real Lorentzian space-time, which is essentially the ``baby universe". Hence the solution of the scalar field wormhole may be interpreted as an ``instanton". Since the variable is cyclic, only the zero mode of the scalar field is imaginary initially, and hence generation of imaginary values of the scalar field in the Lorentzian space does not cause any physical problem. In the present case $\phi$ is cyclic variable and so it should be imaginary in the Euclidean regime.\\

Now since the field equation \eqref{4.2} for $\phi$ reduces to ${d\over dt} (a^3\phi_{,\tau}) = 0$, so we choose

\be\label{4.5} \phi_{,\tau} = i{l\over a^3},\ee
$l$ being a constant of integration. Keeping $\chi$ real, we observe that equation \eqref{4.1} takes the form

\be \label{4.6} a_{,\tau}^2 = 1 - {\alpha\over 6}{l^2\over a^4} +{\beta\over 6} [\chi_{,\tau}^2 -V(\chi)] a^2.\ee
Therefore, our problem essentially reduces to find a solution to equation \eqref{4.3} and \eqref{4.6} under the boundary condition

\begin{subequations}\begin{align}
&  \chi_{,\tau}(0) = a_{,\tau}(0) = 0, \mathrm{and} \\
& \chi(\tau\rightarrow \infty) = \chi_{\infty} = \mathrm{constant,~ together~ with~} a(\tau \rightarrow) = r.\label{4.7}
\end{align}\end{subequations}
For $\chi(\tau) = \chi_{\infty} =$ constant, the solution to equations \eqref{4.3} and \eqref{4.6} reduces to that obtained by Giddings and Strominger \cite{13}. However, even for $\chi(\tau) \ne$ constant, it is possible in principle to find the solutions to these equations under certain simplified assumptions: e.g. equation \eqref{4.3} can be viewed as the motion of a particle in a potential $-V(\chi)$, having coordinate $\chi$, under the influence of a viscous force $3{a_{,\tau}\over a}\chi_{,\tau}$. If the motion of the instanton starts at $(\tau = 0, \chi_0)$, and ends at $(\tau\rightarrow \infty, \chi_{\infty})$, then the frictional term must be negative in some region of $\tau$, and we can consider negligible contributions of the frictional term everywhere except at $\chi_{\infty}$. Under this assumption and in view of equation \eqref{4.3}, the solution to $\chi$ turns out to be

\be \label{4.8} \int_{\chi_0}^{\chi_{\tau}} {d\chi\over \sqrt{2[V(\chi) - V_0]}} = \tau,\ee
where, $V_0$ is a constant of integration. The characteristic time of rolling of the field from $\chi_0$ to $\chi_{\tau}$ is then given approximately by

\be \label{4.9} \Delta\tau = {(\chi_{\infty} - \chi_0)\over \sqrt V_{max}},\ee
$V_{max}$ being the maximum value of the scalar potential, lying somewhere between $\chi_0$  and $\chi_{\infty}$. In the above discussion, $V_{max}$ has been considered to be very large with respect to $V_0$. If the parameters of the present model satisfy certain conditions, then it's possible to show that there are two stages in which the motion of the system proceeds. In the first stage, $\chi$ changes from $\chi_0$ to some point near $\chi_{\infty}$ according to the equation \eqref{4.8}, while $a(\tau)$ becomes approximately equal to $a_0$, which is the value of $``a"$ at $\tau \rightarrow 0$. In the second stage, $\chi(\tau)$ slowly approaches $\chi_{\infty}$, while $a(\tau)$ evolves as the Giddings-Strominger wormhole \cite{13}. The two stages of evolution of $a(\tau)$ are matched at some intermediate value of $\tau$. This wormhole solution corresponds to the creation of an expanding universe. If we analytically continue from Euclidean to the Lorentzian time, then the system will be governed by the equations \eqref{2.8} and \eqref{2.9}. The initial condition at $t = 0$ remains the same as before. The field $\chi$ while moving in the potential $V(\chi)$, oscillates around some point which is damped by positive frictional force. Hence $\chi$ takes a constant value at large $t$. Since $V(\chi) \ge 0$, the created universe expands in an inflationary manner. The inflationary solution in the classically allowed region has been presented in equation \eqref{3.5}, for a slowly varying potential $V(\chi)$. The justification for considering $V(\chi)$ to be a slowly varying function of $\chi$ comes from the above analysis of the wormhole solution.\\

At this point we would like to mention that Rubakov and Tinyakov \cite{20} presented a wormhole solution for a third rank antisymmetric tensor field strength $H_{\mu\nu\lambda}$ coupled with a real scalar field $\phi$ having a potential $V(\phi)$, for a nontrivial function $\phi(\tau)$. It is interesting to note that the conformally invariant Brans-Dicke field and the third rank antisymmetric tensor field are governed by the same type of equations. While Rubakov and Tinyakov model is governed by a single coupling constant $k$ \cite{20}, the present model is governed by a pair viz. $\alpha$ and $\beta$, and this is the essential difference between the two models. The above analysis to obtain wormhole solutions for nontrivial $\chi(\tau)$ is also essentially the same as that of Rubakov and Tinyakov model \cite{20}.\\

So far we have considered the case where only the Brans-Dicke field $\phi$ takes the imaginary value in the Euclidean regime, since it appears in the theory as a cyclic variable corresponding to which there exists a global conserved charge. Recently Burges and Hagan \cite{20} have shown that for a massive scalar field the initial and the final states corresponding to the scalar field prepared in eigenstates of the total energy. This will not be a conserved quantity in general. The interesting feature of the massive scalar field wormhole constructed by Burges and Hagan \cite{20} is that this field is also imaginary in the Euclidean regime, and it correctly describes semiclassical amplitudes between the initial and the final states. In fact, the clain that one would obtain the same Wheeler-DeWitt equation starting from both the Lorentzian and the Euclidean path integral is true only if the Euclidean momentum is imaginary, which is justified if the corresponding field variable is imaginary. In what follows we shall study such wormholes for the present model, where both the cyclic and the non-cyclic variables take the imaginary values in the Euclidean regime. Choosing $\chi = i\theta$ for simplicity, the equation \eqref{4.6} takes the following form

\be \label{4.10} a_{,\tau}^2 = 1 - {\alpha\over 6}{l^2\over a^4} - {\beta\over 6}[\theta_{,\tau}^2 + V(i\theta)] a^2.\ee
Equation \eqref{4.4} now reduces to

\be \label{4.11} a_{,\tau\tau} = {\alpha\over 3}{l^2\over a^5} + {\beta\over 3}[\theta_{,\tau}^2 - V(i\theta)] a.\ee
Substituting the condition for $a_{,\tau} = 0$ from equation \eqref{4.10} in equation \eqref{4.11}, we find that for wormholes to exist i.e. $a_{,\tau\tau} \ge 0$, the following condition has to be satisfied

\be \label{4.12} {\alpha\over 3}{l^2\over a_0^4} + {\beta\over 3} a_0^2\theta_{,\tau}^2 > 1,\ee
where $a_0$ is the radius of the wormhole throat given by

\be \label{4.13} \beta (\theta_{,\tau}^2 + V) a_0^6 - 6 \theta_{,\tau}^2 a_0^2\ge 1.\ee
This equation has two negative roots and one positive. If one considers $\theta_{,\tau}$ to be a constant, Giddings-Strominger wormhole solution \cite{13} is recovered. The general solution to equation \eqref{4.13} for non-trivial $\theta_{,\tau}$ has been discussed by Accetta eta al \cite{21} in a recent publication. However, since wormholes exist at the Planck scale, the radius of the throat $a_0$ is sufficiently small and hence the condition \eqref{4.12} is always satisfied.

\section{Wave Function of the Universe}

This section will be devoted in presenting some solutions to the Wheeler-DeWitt equations \eqref{2.10}. In order to get the wave function of the universe, some boundary conditions should be specified. We shall restrict ourselves to Vilenkin's proposal \cite{17}, which states that the universe spontaneously nucleates from nothing. This boundary condition implies that the wave function $\Psi$ includes only outgoing modes at the singular boundaries of the superspace. Now, since the commutation relation between the scale factor $``a"$ and the conjugate momentum is unknown, factor order ambiguity appears in the Wheeler-DeWitt equation. As noted earlier, $p$ represents some (not all) of the factor ordering ambiguities. Normally, one solves the Wheeler-DeWitt equation with further restriction on $p$. such as choosing a comfortable numerical value of it. The importance of the solutions which we are going to present here is that, at least in the two cases $p$ remains arbitrary within a particular range.\\

\noindent
Case-1\\

The Wheeler-De-Witt equation \eqref{2.10} may be expressed in the following form, if $\Psi$ is a slowly varying function of $\chi$;

\be\label{5.1}  \Psi_{,aa} + {p\over a} \Psi_{,a} - {6\over\alpha a^2}\Psi_{,\phi\phi} -12^2(a^2 - \Lambda a^4)\Psi = 0,\ee
where, $\Lambda = {\beta V(\chi)\over 6}$. Performing the separation of variables $\Psi(a,\phi) = \psi(a)\Phi(\phi)$, one obtains

\be\label{5.2} a^2\left({\psi_{,aa}\over \psi}+{p\over a}{\psi_{,a}\over\psi} \right) -12^2(a^4 - \Lambda a^6) = {6\over\alpha}{\Phi_{,\phi\phi}\over \Phi}={4n^2\over\zeta},\ee
where, both $n$ and $\zeta$ are constants. Therefore.

\be \label{5.3} \Phi_{,\phi\phi}={2\alpha n^2\over 3\zeta},\ee
\be\label{5.4} a^2\left({\psi_{,aa}}+{p\over a}{\psi_{,a}} \right) -\left[{4n^2\over \zeta}+ 12^2(a^4 - \Lambda a^6)\right]\psi = 0.\ee
Let us choose, $a^2 = \zeta y$, so that equation \eqref{5.4} takes the form

\be\label{5.5} \psi_{,yy} + P(y) \psi_{,y} + Q(y)\psi = 0,\ee
\be\label{5.6}\mathrm{where},~~~P(y)= {1+p\over 2y},~~~\mathrm{and},~~~Q(y) = -{n^2\over y^2} - 36\zeta^3(1-4\zeta y).\ee
Now under the choice,

\be\label{5.7} \psi = W(y) \exp{\left(-{1\over 2}\int P dy\right)} = W|y|^{-r}, ~~~\mathrm{where},~~~r = {1+p\over 4},\ee
equation \eqref{5.5} takes the form

\be\label{5.8} W'' +\left(Q - {1\over 2} P' -{1\over 4}|P|^2\right)W = 0,\ee
where prime denotes differentiation with respect to $y$. Further, restricting $r$ to $r^2 -r +n^2 = 0$, i.e. $p = 1+2\sqrt{1-4n^2}$, with $n^2 \le {1\over4}$, equation \eqref{5.8} takes the form

\be \label{5.9} W'' - 36 \zeta^3 (1 - \Lambda \zeta y) W = 0.\ee
Finally, under the choice $1 - \Lambda \zeta y = Z$, the above equation \eqref{5.9} reduces to the following simple form,

\be \label{5.10} W_{,ZZ} - Z W = 0,\ee
where, $\zeta = {\Lambda^2\over 36}$. The solutions to equation \eqref{5.10} are known as Airy functions, whose asymptotic forms are

\begin{subequations}\begin{align}
& A_{iZ\ll 0}(Z) \sim (-\pi^2 Z)^{-{1\over 4}}\sin{\left[{2\over 3}(-Z)^{3\over 2} +{\pi\over 4}\right]}\\
& A_{iZ\gg 0}(Z) \sim {1\over 2}(\pi^2 Z)^{-{1\over 4}}\exp{\left[-{2\over 3}(Z)^{3\over 2}\right]}\\
& B_{iZ\ll 0}(Z) \sim (-\pi^2 Z)^{-{1\over 4}}\cos{\left[{2\over 3}(-Z)^{3\over 2} +{\pi\over 4}\right]}\\
& B_{iZ\gg 0}(Z) \sim (\pi^2 Z)^{-{1\over 4}}\exp{\left[{2\over 3}(Z)^{3\over 2}\right]}\\
\mathrm{where}, ~~~& Z = 1 - \Lambda \zeta y = 1 - \Lambda a^2 = 1- {\beta\over 6} V(\chi) a^2. \label{5.11}
\end{align}\end{subequations}
So, $Z < 0$ corresponds to the classically allowed region, while $Z > 0$ stands for classically forbidden region. The wave function $W$ would be general solution of equation \eqref{5.10} and thus a linear combination of $A_i$ and $B_i$. Thus in general, the solution for the classically allowed region is

\be \label{5.12} W^{\mathrm{osc}} = (-\pi^2 Z)^{-{1\over 4}}\left\{\sin{\left[{2\over 3}(-Z)^{3\over 2} + {\pi\over 4}\right]} +\cos{\left[{2\over 3}(-Z)^{3\over 2} + {\pi\over 4}\right]}\right\},\ee
and the solution for the classically forbidden region is

\be \label{5.13} W^{\mathrm{exp}} = (\pi^2 Z)^{1\over 4}\left\{{1\over 2}\exp{\left[-{2\over 3}(Z)^{3\over 2}\right]} + \exp{\left[{2\over 3}(Z)^{3\over 2}\right]}\right\}.\ee
However, in order to obtain the wavefunction of our universe, one requires only the outgoing modes of the wavefunction to be present. So in the classically allowed region $a^2\Lambda >1$, the tunnelling wavefunction is given by

\be \label{5.14} W_T = {A_i(Z) + iB_i(Z)\over A_i(Z_0) + i B_i(Z_0)},~~~~~Z\ll 0,\ee
where, $Z_0 = Z_{a=0} = 1$. The form of $W_T$ in the classically allowed region and that in the classically forbidden region are essentially the same as the one presented by Vilenkin \cite{17}. Once $W_T$ is known, $\psi_T$ may be obtained and hence the tunnelling wavefunction of our universe is found as $\Psi_T = \Phi \psi_T$, where $\Phi$ is the solution to equation \eqref{5.3}, given by

\be \label{5.15} \Phi = A_1 \exp{\left[2n\sqrt{{\alpha\over 6\zeta}}\phi\right]}+B_1 \exp{\left[-2n{\alpha\over 6\zeta}\phi\right]},\ee
$A_1$ and $B_1$ being the constants of integration.\\

\noindent
Case-2\\

Here, we shall consider the case in the absence of the Higgs field. In view of the separation of variables  $\Psi(a,\phi) = \psi(a)\Phi(\phi)$ as before, the wave equation \eqref{5.1} leads to

\be\label{5,16} {a^2}{\psi_{,aa}\over \psi} + pa{\psi_{,a}\over \psi} - 12 a^4 = {6\over \alpha}{\Phi_{,\phi\phi}\over \Phi} =  - \Omega^2.\ee
Therefore, one obtains the following pair of differential equations, viz.

\begin{subequations}\begin{align}
&\Phi_{,\phi\phi} + {\alpha \Omega^2\over 6}\Phi = 0,\\
& \psi_{,aa} + {p\over a} \psi_{,a} - \left(12 a^2 -{\Omega^2\over a^2}\right)\psi = 0. \label{5.17}
\end{align}\end{subequations}
To solve the equation (54b), we again choose $a^2 = \zeta y$, so that the equation reduces to

\be\label{5.18} \psi_{,yy} +{1+p \over 2y}\psi_{,y} - \left[ 36 \zeta^2 - {\Omega^2 \over 4y^2}\right] \psi =0.\ee

\noindent
\textbf{(a)} \textbf{$p$ is arbitrary}:\\

If we now choose

\be\label{5.19} P(y) = {1+p\over 2y};~~~Q(y) = -36\zeta^2 +{\Omega^2 \over 4y^2};~~~\psi = W|y|^{-r};~~ \mathrm{where},~~r = {1+p\over 4},\ee
then equation \eqref{5.18} takes the form

\be \label{5.20} W'' +\left[-36\zeta^2 +{{\Omega^2\over 4} +r -r^2\over y^2}\right] W = 0.\ee
Further, under the choice ${\Omega^2\over 4} +r -r^2 = {1\over 4} - n^2$, $\zeta = {1\over 6}$ and $z = -iy$, the above equation \eqref{5.20} reduces to

\be \label{5.21} W_{,zz} + \left(1 - {n^2 - {1\over 4}\over z^2}\right)W = 0,\ee
the general solution to which is

\be \label{5.22} W_n = \sqrt{z} [A J_n(z) + B J_{-n} (z)],\ee
where $n$ is not an integer and $J_n$ is the Bessel function of first kind. In terms of the variable $y$, the solution takes the form

\be\label{5.23} W_n = \sqrt{y} \left\{(A){\pi\over 2}\left[{I_{-n}(y) - I_n (y)\over \sin{n\pi}}\right]\right\},\ee
for all values of $n$. Hence,

\be \label{5.24} \psi_n = (6a^2)^{1-2r\over r} \left\{{(A)\pi\over 2}[I_{-n}(6a^2) - I_n (6a^2)]\right\}.\ee
The solution to the equation (54a) is

\be \label{5.25} \Phi = A_1 \exp{\left(\sqrt{\alpha\over 6} \Omega \phi\right)} + A_2 \exp{\left(-\sqrt{\alpha\over 6} \Omega \phi\right)},\ee
where, $A_1$ and $A_2$ are constants of integration. Hence the solution to the wavefunction $\Psi(a, \phi)$ may be obtained independent of the choice of the factor ordering parameter $p$.\\

\noindent
\textbf{(b) $p$ = 1}:\\

For the particular choice of the factor ordering parameter $p = 1$, equation \eqref{5.18} reduces to

\be \label{5.26} y^2 \psi_{,yy} + y\psi_{,y} - \left(y^2 - {\Omega^2\over 4}\right) = 0.\ee
Now, under the choice $\zeta = {1 \over 6}$, ${\Omega^2\over 4} = - n^2 > 0$, the above equation \eqref{5.26} reduces to

\be\label{5.27} y^2 \psi_{,yy} + y\psi_{,y} - \left(y^2 + n^2\right) = 0.\ee
The solution to equation \eqref{5.27} is given by the modified Bessel function in the following form

\be \label{5.28} \psi_n = A I_n(y) + BI_{-n} (y),\ee
for arbitrary values of $n$, which is not an integer. However, for integral values of $n$, $I_n(y) = I_{-n}(y)$, and the solution may then be represented by the modified Bessel function of second kind

\be \label{5.29} \psi_n = k_n(y) = {\pi\over 2}\left[{I_{-n}(y) - I_n (y)\over \sin{n\pi}}\right].\ee
The above solution of $\psi_n$ is well behaved for all values of $n$. Particulary, in the asymptotic limit $y \rightarrow \infty$, the solution reduces to

\be \label{5.30} \psi_n \rightarrow \sqrt{\pi\over 12} \left[{\exp{(-6a^2)}\over a}\right].\ee

\noindent
\textbf{(c) $p = 3$}\\

For $p = 3$, under the same above choice $\zeta = {1\over 6}$ and ${\Omega^2\over 4} = - n^2, ~n^2> 0$, equation \eqref{5.18} takes the reduced form

\be \label{5.31} y^2 \psi_{,yy} + 2y\psi_{,y} - \left(y^2 + n^2\right)\psi = 0.\ee
The above equation admits a general solution in the form of spherical modified Bessel function, viz.

\be \label{5.32} \psi_n = \sqrt{\pi\over 2y}~I_{n+{1\over 2}}(y).\ee
It is important to mention that, as $y \rightarrow 0$, $I_n(y)$ behaves as the Bessel function $J_n(y) \sim {1\over \Gamma(n+1)}\left({y\over 2}\right)^n$.\\

\noindent
\textbf{(d) $p = -1$}\\

This choice of operator ordering index was considered by Vilenkin \cite{17}. Under the choice $\Omega^2 = 1- 4m^2$, $\psi = \eta\sqrt y$, $\zeta = {1\over 6}$ and $a^2 = \zeta y$, equation \eqref{5.18} reduces to the following form

\be \label{5.33} \eta_{yy} + {\eta_y\over y} - \left(1+{m^2\over y^2}\right)\eta = 0.\ee
The solution to the above equation after changing the variable to $\psi$ takes the form

\be \label{5.34} \psi_m = \sqrt{6} {\pi a\over 2\sin{m\pi}}[I_{-m}(6a^2) - I_m (6a^2)].\ee
In the asymptotic limit $a \rightarrow \infty$, the wavefunction $\psi_m \rightarrow \exp{(-6 a^2)}$, while as $a \rightarrow 0$, it vanishes, i.e. $\psi_m \rightarrow 0$.

\section{Conclusion}

We have explored different classical and quantum-cosmological aspects of Brans-Dicke scalar-tensor theory coupled with a Higgs field. The important results that have been obtained are: (i) In general Brans-Dicke field does not allow a graceful exit from inflation; and (ii) the wave function of the universe can be obtained even without fixing the factor ordering index. In addition, we have presented viable wormhole solutions with and without conserved charges.

\end{document}